\documentclass[amsmath,amssymb,onecolumn,showpacs]{revtex4}
\usepackage{amsfonts}
\usepackage{mathrsfs}
\usepackage{graphicx}  
\usepackage{cases}

\begin{document}

\title{Quench of non-Markovian coherence in the deep sub-Ohmic spin-boson model: A unitary equilibration scheme}

\author{Yao Yao}
\email[Email: ] {yaoyao@fudan.edu.cn}
\affiliation{State
Key Laboratory of Surface Physics and Department of Physics, Fudan
University, Shanghai 200433, China}

\date{\today}

\begin{abstract}
The deep sub-Ohmic spin-boson model shows a longstanding
non-Markovian coherence at low temperature. Motivating to
quench this robust coherence, the thermal effect is unitarily
incorporated into the time evolution of the model, which is
calculated by the adaptive time-dependent density matrix
renormalization group algorithm combined with the orthogonal
polynomials theory. Via introducing a unitary heating operator to
the bosonic bath, the bath is heated up so that a majority portion
of the bosonic excited states is occupied. It is found in this
situation the coherence of the spin is quickly quenched even in the
coherent regime, in which the non-Markovian feature dominates. With
this finding we come up with a novel way to implement the unitary
equilibration, the essential term of the eigenstate-thermalization
hypothesis, through a short-time evolution of the model.

\end{abstract}


\maketitle

\section{introduction}

The molecular materials, both organic and biological, are usually
with high degree of dynamic disorder, which induces the strong
effect of localization and quantum decoherence. It then seems
reasonable to account the intrinsic mechanism of the charge
conduction incoherent. However, as widely accepted the transport
mechanism in these materials undergoes a transition from coherent to
incoherent at finite temperature \cite{review0}. Many recent
experiments then devoted to uncover the coherent component
accordingly \cite{Friend1,Friend2,Friend3,Guo}. For example, the
transient absorption spectrum experiment shows an ultrafast charge
transfer process following with a long-termly incoherent decay of
the population of the induced charge transfer state \cite{Friend1}.
These advances of the experiment are deserving the appropriate
theoretical explanations, but the present theoretical study lags
behind. The significant issue to hinder the progress is that, the
decoherence process is not easy to be self-consistently involved in
the coherently quantum dynamics \cite{deco}. For example, in the
dynamic disorder model we have shown that the electron's coherence
can not be quenched by the motion of lattice without any artificial
implements of decoherence \cite{mine2}. Consequently, we are on the
stage of thinking about how to smoothly adapt the decoherence into
the pure-state quantum dynamics and incorporate the two kinds of
mechanisms, coherent and incoherent, into a unified framework.

Our motivation of the present work is based upon the progress of the
thermalization emerged in the last several years, which is achieved
through a purely quantum-mechanical manner
\cite{ther0,ther1,ther2,ther3,ther4,ther5,ther6}. Of the most
importance in the field is the eigenstate-thermalization hypothesis
(ETH), which states that while evolving, a random initial
\textit{pure state} of the universe, namely the composite of a
system plus a bath, produces the same thermal average of the
observables of the system \cite{ther2,ther3}. Moreover, another
significant statement addresses that a sufficiently small system
weakly coupled with a bath always thermalizes no matter how far it
is initially from thermal equilibrium \cite{ther4}. These novel
concepts are moving the understanding of the quantum statistical
mechanics forward presently. In this context, reexamining the
statement in different system-bath coupling models goes active and
comprehensive
\cite{ther7,ther8,ther19,ther9,ther10,ther11,ther12,ther13,ther14,ther15,ther16,ther17,ther18}.
Initially, the adaptive time-dependent density matrix
renormalization group (t-DMRG) studies of both one-dimensional
spinless fermion and bose-Hubbard model after a quantum quench
showed the breakdown of thermalization for both integrable and
nonintegrable cases \cite{ther5,ther6}. Soonafter, the exact
diagonalization studies gave distinct results for integrable and
nonintegrable cases of both one-dimensional boson and fermion models
\cite{ther7}. Following this line, the thermalization and ergodicity
are both demonstrated for one-dimensional open systems
\cite{ther1,ther8,ther9}.

By definition, if the ETH holds, the thermalization of an observable
of the system is guaranteed for any pure state of the composite with
well-defined macroscopic observables \cite{ther0,ther1}. There are
three elements to describe this so-called quantum thermalization
\cite{ther4}. (1) If the system (under the influence of bath)
initially far from equilibrium always evolves to and remains in a
particular state, we call it equilibration. (2) The equilibrated
state of the system should be independent of the microscopic details
of the initial states of both the system and the bath. (3) In a
certain degree, the population of equilibrated state should take the
form of Boltzmann distribution. The former two terms are attributed
to be generic for the quantum thermalization, while the third one
calls for a specific coupling term in the Hamiltonian. Our present
work is motivated to incorporate the decoherence into the pure-state
dynamics, namely to properly involve the influence of bath in the
time evolution of the system. So hereafter, we will mainly focus on
the equilibration of the system, that is the first two terms of the
quantum thermalization.

It has been addressed that the necessary condition of quantum
thermalization relies on whether the available space of the bath is
large enough such that the coherence of system can be easily
quenched \cite{ther4}. Or alternatively speaking, whether the system
thermalizes strongly depends on the \textit{decoherence} induced by
the bath \cite{ther17,ther18}. So far, several system-bath coupling
models have been studied comprehensively for the subject of
thermalization. For example, a nine-site Hubbard model, which is
numerically exactly solvable, was investigated by considering two
sites out of the nine to be the system \cite{ther18}. Here, as we
are interested in the transportation mechanisms in molecular
materials, the spin-boson model (SBM) will be taken into account. To
our knowledge, the study of the quantum thermalization of this model
is still absent in the researches up to now. The Hamiltonian of SBM
reads \cite{review1,review2}, ($\hbar\equiv1$)
\begin{eqnarray}
H=H_s+\frac{\sigma^z}{2}\sum_l\lambda_l(b^{\dag}_l+b_l)+\sum_l\omega_l
b^{\dag}_lb_l,\label{hami}
\end{eqnarray}
where $H_s=\frac{\varepsilon}{2}\sigma^z-\frac{\Delta}{2}\sigma^x$
is the Hamiltonian for the spin with $\sigma^z$ and $\sigma^x$ being
the usual notations for Pauli operators, $\varepsilon$ the bias
induced by the external magnetic field, and $\Delta$ the tunneling
constant; $\lambda_l$ is the coupling constant between the spin and
the boson of $l-$th mode, $\omega_l$ the respective frequency, and
$b^{\dag}_l (b_l)$ the creation (annihilation) operator of bosons.
The degree of freedom of the spin is much smaller than that of
bosons, so we can safely recognize the spin as a small system and
the bosons as the bath. In addition, the SBM is a highly abstract
model to simulate the transportation in molecular materials, such as
the excitation energy transfer process \cite{Excite}. The spin
represents a two-level system and the bosons simulate the phonon
environment. So in principle our study here could be helpful for the
understanding of coherent-incoherent transition of transportation in
the relevant materials.

The frequency of the boson in SBM is usually cut off at $\omega_c$,
which is chosen to be 1 as unit. The spectral density function of
the bath is expressed as
$J(\omega)=2\pi\alpha\omega^{1-s}_c\omega^s$ with $\alpha$ being the
dimensionless spin-boson coupling. The regime $s<1$ refers to the
sub-Ohmic bosonic bath showing the strong non-Markovian feature,
which is addressed to induce long-term coherence and dominate the
dynamics \cite{SBM3}. In particular, very recent studies indicated
that \cite{SBM1,SBM2}, at low temperature there is a persistently
coherent regime for deep sub-Ohmic case ($s<0.5$) so that the spin
keeps oscillating during the time evolution. In order to simulate
the coherence in the transportation of molecular materials, we will
then work in this regime in the present work. Especially, as the
robust non-Markovian featured coherence in this regime emerges
contradictorily to the decoherence requirement of ETH, it is then
available to study the situation that the dynamics of the system
loses the coherence (or reaches equilibrated state equivalently)
during a \textit{short-time} quantum evolution.

Consequently, ETH is claimed to be as generic as the concept of
thermalization, but the studies of the pure-state dynamics of deep
sub-Ohmic SBM shows contradictions to it. In order to solve this
seemingly paradoxical problem, in this work, we investigate the
quantum dynamics of the model in the coherent regime following the
first two requirements of ETH. A steady value of spin population,
namely the equilibration, is obtained and the non-Markovian feature
is shown to be quickly quenched. The paper is organized as follows.
The methodology we use is introduced in Sec. II. Calculation results
for both one spin and two spins are presented in Sec. III, where the
thermalization and effective temperature are discussed. Conclusions
are drawn in the final section.

\section{Methodology}

As mentioned above, the necessary condition of thermalization in a
quantum-mechanical fashion is, in our opinion, the decoherence time
of the system is shorter than the relaxation time. In this situation
the quantum coherence of the system could be quickly quenched before
the thermalization is achieved. This condition requires that a
sufficiently large eigenstate space of the bath has been (or at
least could be) occupied \cite{ther4}. However, this is not the case
in SBM especially in the coherent regime. Firstly, the strong
non-Markovian feature of the bath always gives rise to the
longstanding coherence and memory of the initial state. Secondly,
based upon the variational theory, during the time evolution, the
sub-Ohmic bath could be perfectly described by some combinations of
the coherent states \cite{mine}. The number of these states is not
large enough to quench the coherence. Therefore, it is necessary to
find a novel way to enlarge the number of available states of the
bath.

To this end, we first make a transformation to the Hamiltonian. In
the present form of Hamiltonian (\ref{hami}), each bosonic mode of
the bath has its own channel to influence the spin, and the coupling
in between cannot be regarded to be relatively weak. Alternatively,
the theory of orthogonal polynomials presents one way to solve this
problem \cite{Chin}, \textit{i.e.}, to transform the bosonic modes
into a one-dimensional chain, with each site in the chain
representing a renormalized mode, and most importantly, the spin
interacts with only the first site of the chain. This transformation
is exact, via which we are able to make the spin-bath coupling
change from nonlocal to local and minimize the action of the bath on
the spin. It then becomes available to consider a really ``small"
system (the spin) embedding in a ``large" bath (the bosonic chain).
The transformed Hamiltonian is written as
\begin{eqnarray}
\tilde{H}&=&H_s+\sqrt{\frac{\eta}{4\pi}}\sigma^z(b^{\dag}_1+b_1)+\sum_n\omega_nb^{\dag}_nb_n\nonumber\\&+&\sum_n(t_nb^{\dag}_{n+1}b_n+{\rm
h.c.}),\label{hami2}
\end{eqnarray}
where $\eta$ is the renormalized spin-boson coupling, which could be
estimated by $\eta=\int_0^{\omega_c}J(\omega)d\omega$; $n$
represents the $n-$th renormalized mode (bosonic site), and
$\omega_n$ and $t_n$ are the respective transformed frequency and
hopping integral among $n$ and $n+1$ site, whose precise expressions
could be found in Ref. \cite{Chin,mine}. On each site, the dimension
of the Fock space must be cut off at some finite number. Our recent
work has tested different parameters and compared the results with
that from two other approaches \cite{mine}. The comparison shows
that, the results are reliable when the dimension of the Fock space
of each site is cut off at $4$. Here we will keep working with the
same parameter to make sure the results are credible.

\begin{figure}
\includegraphics[angle=0,scale=0.43]{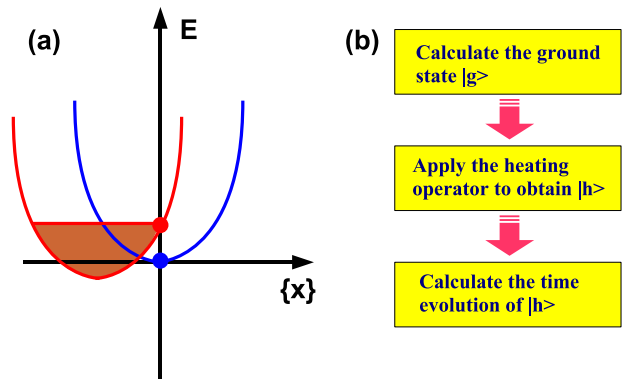}
\caption{(a) Schematic for the effect of heating operator (see
definition in the text). Initially, the bath is around the bottom of
the potential function (blue point). With the action of heating
operator, the potential energy changes, but the configuration of
displacement keeps unchange (red point). Hence, the possible number
of states that the bath may occupy increases. (b) The flowchart of
the calculating procedure with t-DMRG algorithm. $|g\rangle$ and
$|h\rangle$ denote the ground state and heated state,
respectively.}\label{sch}
\end{figure}

The dynamics based upon Hamiltonian (\ref{hami2}) could be
calculated by the adaptive time-dependent density matrix
renormalization group (t-DMRG) algorithm \cite{mine}. The spin and
the bosonic sites together form a one-dimensional lattice chain with
only nearest-neighbor interaction. The topology of this lattice is
friendly for DMRG calculation and we have obtained precise results
with DMRG truncating number of $100$ \cite{mine}. The calculating
procedure in this work is divided into three steps. (i) We set
$\varepsilon$ to $-0.5$ and calculate the ground state $|g\rangle$
of the Hamiltonian (\ref{hami2}). This will make the spin freeze in
its up state and the bath almost in its vacuum state. (ii) Based on
the calculated ground state $|g\rangle$, the bath is heated up under
the action of the operator $\mathscr{H}$ described in the next
paragraph. (iii) We change $\varepsilon$ to the desired value and
calculate the time evolution of the whole system with the initial
state being $|h\rangle=\mathscr{H}|g\rangle$. The whole procedure is
sketched in Fig.~\ref{sch}(b) and is based on the concept of
pure-state evolution, which is consistent with the premise of ETH.

The central point in the calculating procedure is to appropriately
introduce the thermal effect into the bath, i.e., to let the bath
initially be with some kind of heated state. Although there have
been some temperature-dependent DMRG algorithms \cite{tempDMRG},
they are both inefficient and impractical for the present model. To
this end, we first introduce an operator of the bath, i.e.,
$\mathscr{D}=\mathcal {\delta}\sum_n(b^{\dag}_n+b_n)$, with
$\mathcal{\delta}$ being an effective polarization field for each
mode. The motivation of this introduction is as follows. As the
operator $\mathscr{D}$ commutes with the coupling term of the
Hamiltonian, namely the second term of the right hand side of
(\ref{hami2}), the action of it does not change the coupling energy
with respect to the coupling term. Its effect is only to enhance the
energy of the bath itself, which is what we want. Furthermore, in
order to minimize the possible error that the operator $\mathscr{D}$
may bring to the computations, we introduce another operator
$\mathscr{H}\equiv e^{-i\mathscr{D}t_H}$ as mentioned above, with
$t_H$ the action time of the polarization field onto the bath. This
operator is unitary so that we can divide $t_H$ into many time steps
and apply $\mathscr{H}$ to the state calculated by DMRG. The
precision of the action of the unitary operator $\mathscr{H}$ could
be well controlled and the results are then highly reliable.
Subsequently, we have introduced two parameters $\mathcal{\delta}$
and $t_H$. For simplicity, in the practical computations
$\mathcal{\delta}$ is set to be unity and the final results are
dependent of $t_H$.

To make it clear, Fig.~\ref{sch}(a) shows a schematic of the effect
of the introduced operator $\mathscr{H}$. Suppose the bath initially
stays at the vacuum state. As the operator $\mathscr{H}$ is acted on
the bath itself, the kinetic energy of the bath, or equivalently the
internal energy, could be largely increased by its action. In this
situation much more bath states could be occupied than the case that
the bath is at the bottom of the energy potential. The essential
points of the advantage of this adaption are then worth noting.
Firstly, as the operator $\mathscr{H}$ does not commute with the
momentum operator of bosons, the effect of the operator is to
effectively change the kinetic energy of the bosons which makes the
motion of bosons dynamically disordered. This effect is equivalent
to heating up the bath in a duration of $t_H$, so the operator
$\mathscr{H}$ could be regarded as a ``heating" operator and $t_H$
the ``heating" time. Secondly, as the initial state of the time
evolution is a pure state, the unitary heating operator
$\mathscr{H}$ will keep the system-bath composite in a pure state.
Hence, we are always working with the pure state instead of some
mixed state as in many other algorithms, such as the Lindblad master
equations \cite{ther9}, the Monte-Carlo based approach \cite{QUAPI}
and the polaron theory \cite{Zheng}.

\section{Results and discussions}

In this section, we will mainly discuss our calculating results for
the thermalization of deep sub-Ohmic SBM. In particular, as the
thermalization proves for the relatively weak coupling case,
throughout this work we will focus on the regime of $s=0.25$ and
$\alpha\leq0.03$ in which the non-Markovian feature is very robust
at low temperature.

\subsection{Unitary equilibration of the spin population}

\begin{figure}
\includegraphics[angle=0,scale=2.6]{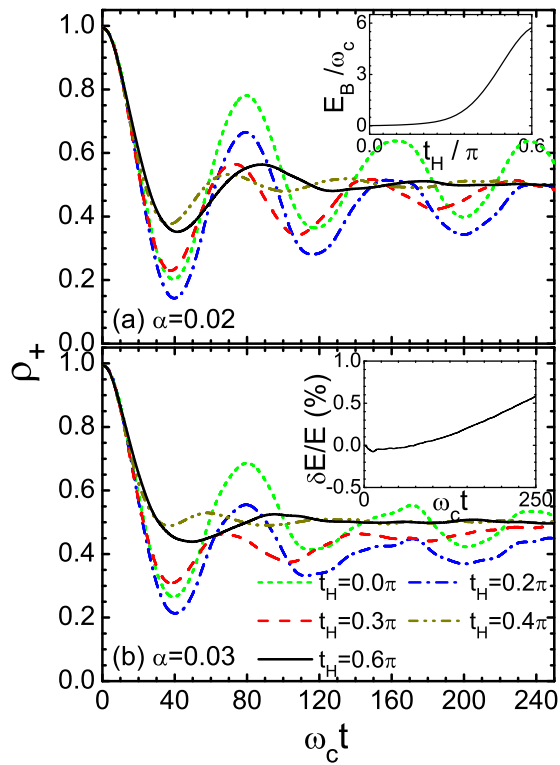}
\caption{Time evolution of up-state population for five heating
times $t_H$ with (a) $\alpha=0.02$ and (b) $\alpha=0.03$. Inset of
(a) shows the dependence of bath energy on $t_H$. Inset of (b) shows
the relative deviation of the total energy during the time
evolution. The other parameters are: $s=0.25, \Delta=0.1,
\varepsilon=0$.}\label{tH}
\end{figure}

We first show in Fig.~\ref{tH} the dependence of the spin's up-state
population evolution on $t_H$ for $\Delta=0.1, \varepsilon=0$ and
$\alpha=0.02$ and $0.03$.  It is found that when $t_H$ is relatively
small, namely $0.2\pi$, the oscillating behavior of the spin is very
similar to that at zero temperature ($t_H=0$), regardless of the
coupling strength. This oscillation is dominated by the
non-Markovian feature of the bath, as frequently discussed in the
literature \cite{mine}. There is a small shift of the amplitude of
spin population in between, which is very similar with that of
changing the initial state from factorized to polarized \cite{SBM3}.
This comparison indicates that when $t_H$ is small our heating
procedure is actually to make an initial displacement of the bath.

Then we increase $t_H$ and the situation changes. The oscillation is
quickly quenched after several periods, \textit{e.g.} three periods
for $t_H=0.3\pi$, two periods for $t_H=0.4\pi$, and one period for
$t_H=0.6\pi$. Especially, when $t_H$ is increased to $0.6\pi$, the
up-state population quickly evolves towards a steady value just
after one oscillation period. According to the ETH, this is the
so-called equilibration. As we have stated, since the heating
operator $\mathscr{H}$ is unitary, the state of the system-bath
composite is always a pure state, and due to the strong
non-Markovian effect the dynamics of a pure state is traditionally
expected to be sensitive to the initial state and keep oscillating
for a long time duration \cite{SBM3}. Our results clearly show a way
to break this consequence and to quench the non-Markovian feature.
To our knowledge, this effect has not been obtained in the coherent
regime through a short-time evolution without any non-unitary
adaption. Our present results then establish a prototype for the
quench of non-Markovian featured coherence in a unitary fashion,
which is the essential result of the present work.

In the inset of Fig.~\ref{tH}(a), we show the $t_H$ dependence of
the bath's energy with respect to the last two terms of Hamiltonian
(\ref{hami2}). $t_H$ is chosen to be smaller than $0.6\pi$ since our
numerical method has truncated the Fock space of bosons and restrict
the energy in an extent smaller than about $6\omega_c$. In the inset
of Fig.~\ref{tH}(b), the relative deviation of the total energy
during the time evolution is shown. Ideally it should be vanishing
due to the unitarity of our calculating procedure. Here, we find the
deviation is always smaller than $1\%$ which is an acceptable
precision for the numerical method.

\begin{figure}
\includegraphics[angle=0,scale=0.6]{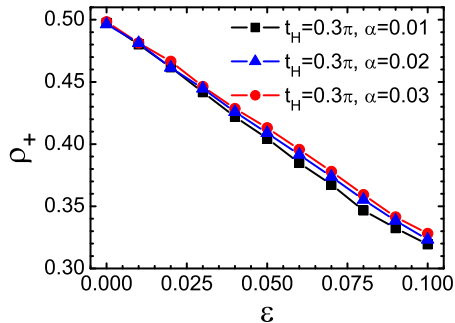}
\caption{Equilibrated value of up-state population versus
$\varepsilon$ for three $\alpha$'s with $t_H=0.3\pi$. The other
parameters are: $s=0.25, \Delta=0.1$.}\label{Bo}
\end{figure}

In Fig.~\ref{Bo}, we show the $\varepsilon$ dependence of
equilibrated value of the up-state population for $t_H=0.3\pi$. The
equilibrated value is obtained when $\rho_{+}$ becomes unchanged
after a long time duration (about $300\omega_c^{-1}$ for
$t_H=0.3\pi$). For $\varepsilon=0$, the equilibrated value of
up-state population is $0.5$, the expected thermalized value
according to the Boltzmann distribution. Following $\varepsilon$
increases, the equilibrated value of the up-state population goes
down, implying the population of higher-energy state decreases. This
finding suggests the equilibration has appropriately taken place as
expected. Meanwhile, we can find from Fig.~\ref{Bo} that the
coupling constant $\alpha$ almost does not influence the
equilibrated value, as for the three $\alpha$'s we choose the change
of the equilibrated population is less than $0.01$ within the extent
of numerical error. This provides another evidence for the
thermalization that the equilibration does not depend on the
coupling between system and bath.

\begin{figure}
\includegraphics[angle=0,scale=1.6]{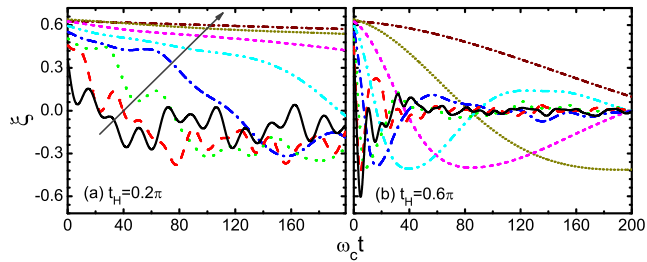}
\caption{Time evolution of $\xi$ on odd sites of boson mode for (a)
$t_H=0.2\pi$ and (b) $t_H=0.6\pi$. The arrow indicates the direction
of site index increasing. The other parameters are: $s=0.25,
\Delta=0.1, \alpha=0.02$. }\label{xi}
\end{figure}


So far, one would be wondering why our method is able to produce the
equilibration in a completely unitary manner. Or one may ask what is
the difference between the heated state $|h\rangle$ that our method
prepares (for large $t_H$) and the usual initial states, such as the
polarized state \cite{SBM3} in which each bosonic mode is driven to
get an individual initial displacement. In Fig.~\ref{xi}, we show
the comparison of the displacement dynamics for bosonic sites
between $t_H=0.2\pi$ and $0.6\pi$. Here, $\xi$ is defined as
$\xi\equiv\langle \hat{b}^{\dag}+\hat{b}\rangle/2$. It is shown
that, when $t_H$ is small, there are several bosonic sites, whose
displacement does not change very much during the time evolution. As
we have discussed above, this is very similar with the case of
polarized initial state. The reason of the effect is that, the
hopping integral in Hamiltonian (\ref{hami2}) decreases with respect
to the distance from the spin and those sites far away from the spin
hardly participate in the dynamics. On the other hand, however, when
$t_H$ is large, more sites are active due to the initial action of
heating operator. This comparison tells us that because of the
participation of more bosonic sites, much more bosonic states than
that in the usual approaches begin to play significant roles in the
dynamics, such that the spin dynamics can be damped and the
thermalization takes place. This improvement benefits from the
action of heating operator, which acts on every bosonic sites and
make them working. Consequently, we would like to address that, this
approach of activating the bosonic states as many as possible could
be generalized to the quantum-classical methods to self-consistently
study the transition from coherent to incoherent mechanism.

\subsection{Effective temperature}

\begin{figure}
\includegraphics[angle=0,scale=0.6]{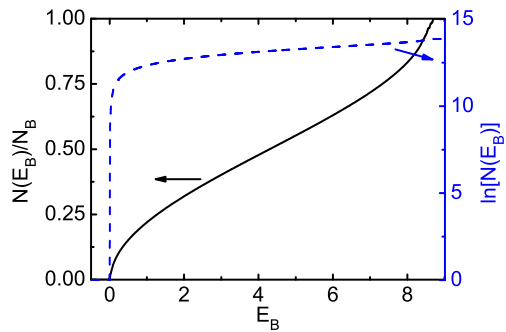}
\caption{The number of states $N(E_B)$ with linear and logarithmic
scale (black solid and blue dashed line) versus the energy of bath
$E_B$. Herein, since $N(E_B)$ is an extremely large number, we make
$N_B$, the total number of the bath states, as the unit.
}\label{tem}
\end{figure}

As to a generic definition of the thermalization, it is necessary to
define a temperature to make sense of the thermal distribution. In a
common sense, the usual definition of an effective temperature for
the quantum models relies on a sufficiently long-time evolution,
during which the system is always expected to evolve to a thermally
equilibrated state. But in this work, the heated bath quenches the
robust non-Markovian feature and quantum coherence during a
short-time duration. In this context one would then wonder whether
the heating effect obtained here is equivalent to that based upon
the usual definition of temperature. In a generic manner they are
not necessarily to be the same, since as mentioned the Boltzmann
distribution calls for the specific coupling term in the Hamiltonian
and needs to be studied case by case. So in the following let us
discuss the present SBM specifically.

In our numerical method, the number of Fock states on each site of
the bath has been truncated to $4$, so the number of the available
bath states in the evolution could be computed although it is
extremely large. Fig.~\ref{tem} shows the relationship between the
number of states and the energy of the bath. From this figure, the
effective temperature of the bath could be calculated by
$T=(\partial\ln N(E_B)/\partial E_B)^{-1}$ with $k_B=1$, $E_B$ being
the energy of the bath with respect to the last two terms of
Hamiltonian (\ref{hami2}), and $N(E_B)$ being the number of bath
states with energy smaller than $E_B$. Here, $N(E_B)$ is calculated
by summing up the number of states from $0$ to $E_B$. It is
cumulative since we realize that all the bath states with energy
below $E_B$ are active and playing roles in the evolution. Hence, by
heating the bath to the corresponding energy, we then get a bath
with the effective temperature $T$ defined here. The inset of
Fig.~\ref{tH}(a) has shown the relationship between the energy of
the bath $E_B$ and the heating time $t_H$. Combined with
Fig.~\ref{tem}, the relationship between $t_H$ and the effective
temperature $T$ could then be established quantitatively. In
particular, for $s=0.25$ and $\alpha=0.02$, the value $t_H=0.2\pi$
is found to be equivalent to the effective temperature
$T\simeq0.1\omega_c$, $t_H=0.3\pi$ to $T\simeq0.6\omega_c$,
$t_H=0.4\pi$ to $T\simeq2.5\omega_c$ and $t_H=0.6\pi$ to
$T\simeq5.6\omega_c$.

Before ending this subsection, we discuss more about the thermal
distribution. For the Boltzmann distribution function we can define
a formula $\exp(-W/2T)$ with $W=\sqrt{\varepsilon^2+\Delta^2}$ being
the energy difference between the spin's eigen-energies depending on
both the $\varepsilon$ and $\Delta$. But in the SBM we study the
coupling term involves the operator $\sigma^z$, namely the preferred
spin states of the bath are almost the up- and down-state on
\textit{z} orientation. This implies that the equilibrated spin
population might be more dependent of the bias $\varepsilon$, so the
equilibrated up-state population versus the $\varepsilon$ shown in
Fig.~\ref{Bo} could be a good reference of the validness of thermal
distribution in the SBM. However, a rigorous verification needs a
lot of computations which is out of the scope of the present work.

\subsection{Quench of entanglement}

\begin{figure}
\includegraphics[angle=0,scale=0.6]{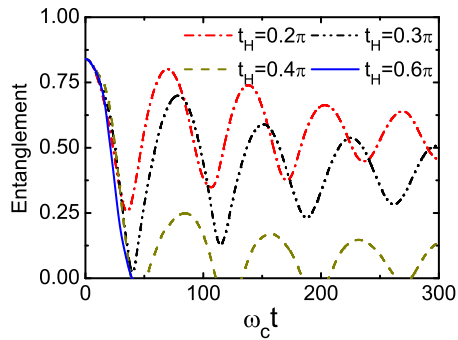}
\caption{Time evolution of the entanglement between two spins for
four $t_H$'s with $\alpha=0.02$. The other parameters are: $s=0.25,
\Delta=0.05, J=0.1$. }\label{Ent}
\end{figure}

To further convince the unitary equilibration we should compare our
results with that from the mixed-state based approach. It has been
found qualitatively by the benchmarking quasiadiabatic propagator
path integral (QUAPI) \cite{QUAPI} that at finite temperature the
quantum entanglement between two spins in a common bath induced by
the non-Markovian feature will be completely quenched. In order to
see whether our pure-state evolution could give rise to the similar
effect, it is necessary to involve two spins $1$ and $2$ in the
Hamiltonian (\ref{hami2}). Namely, the Hamiltonian of spin $H_s$ is
replaced by
\begin{eqnarray}
H'_s=\sum_{\mu=1,2}\frac{\epsilon}{2}\sigma_{\mu}^z-\sum_{\mu=1,2}\frac{\Delta}{2}\sigma_{\mu}^x+J\boldsymbol{\sigma}_1\cdot\boldsymbol{\sigma}_2,
\end{eqnarray}
where $J$ being the exchange constant between the spins and
$\boldsymbol{\sigma}_{1,2}$ the Pauli operator for spin 1 and 2,
respectively. The coupling terms between the respective spin with
the common bath take the same formula as that in (\ref{hami2}). The
parameters in the new Hamiltonian are chosen as $\epsilon=0,
\Delta=0.05$ and $J=0.1$ such that the system is initially in an
entangled state. The exchange constant $J$ is large enough to induce
a strong entanglement between the two spins. Then the bath is heated
by the heating operator $\mathscr{H}$ via the same procedure
described above. Starting with the heated state we calculate the
time evolution of the entanglement between the two spins which is
measured by the concurrence. The calculating procedure of the time
evolution is the same with that in our preceding work \cite{mine}.

In Fig.~\ref{Ent}, we show the results for various $t_H$ and
$\alpha=0.02$. We can find that when $t_H$ is small, namely the
bath's energy is low, the entanglement evolution shows an
oscillating behavior for a long duration. Following the $t_H$
increases the entanglement decreases and when $t_H=0.4\pi$ the
effect of entanglement sudden death and revival appears. We further
increase the $t_H$ to $0.6\pi$, the entanglement is completely
quenched in a very short duration ($\omega_ct<50$). In a
quantum-mechanical manner, the oscillation, sudden death and revival
of entanglement are clearly signatures of the non-Markovian
behavior. With increasing $t_H$ these features are gradually
suppressed and finally completely quenched. Notice that different
from the quench of the oscillation behavior for one-spin case, the
entanglement here is not completely quenched within the duration
$\omega_ct<300$ for $t_H=0.3\pi$ and $0.4\pi$. This is simply
because the time scale of the two-spin system is different from that
of one-spin system. Subsequently, we would like to indicate that the
quench of the non-Markovian features in two-spin system is still
closely related to the unitary equilibration as we discussed above.
It also helps us to make sense of the unitary equilibration in the
non-Markovian dominated SBM.

\section{Conclusion}

In summary, we have investigated the thermodynamics of the deep
sub-Ohmic spin-boson model. A heating operator is introduced, whose
action is to enhance the energy of the bosonic bath without changing
the coupling energy. Via this operation, the bath is heated up to an
effective temperature and the dynamics of the spin embedding in the
heated bath is intensively discussed. The equilibration is found to
take place regardless of the coupling strength and the non-Markovian
behavior induced by the bath is shown to be quickly quenched.
Consequently, we have presented an applicable way to study the
unitary equilibration of the deep sub-Ohmic SBM. Especially, our
study suggests a novel way to quench the quantum coherence in the
dynamics in molecular materials, that is, to let as many as possible
bath states to participate in the time evolution. This conclusion is
generalizable for the subjects of coherent-incoherent transition. In
addition, the method we develop could also be applied to many other
related subjects, such as the heat current transmission through a
spin sandwiched in two heated bosonic baths.

\begin{acknowledgments}
This work was supported by the National Natural Science Foundation
of China (91333202, 11134002 and 11104035), and the National Basic
Research Program of China (2012CB921401). The author thanks C. Q.
Wu, W. Yang, Z. L\"{u}, and Y. Zhao for helpful discussions and
comments on the manuscript.
\end{acknowledgments}

\end{document}